%% file: main.tex
\documentclass[sigconf]{acmart}

\usepackage{bm}
\usepackage{listings}
\usepackage{multirow}
\usepackage{hyperref}
\usepackage{booktabs}
\usepackage{multirow}
\usepackage{siunitx}

\AtBeginMaketitle{%
  \input{edbt-macros-v29-n3}
}

 \AtBeginDocument{%
   }

 \newcommand{\stitle}[1]{\bigskip\noindent\textbf{#1}}

  \hypersetup{
     colorlinks,%
     citecolor=black,%
     filecolor=black,%
     linkcolor=black,%
     urlcolor=black,%
     pdftitle={Data-driven Trajectory Imputation for Advanced Vessel Monitoring},
     pdfauthor={Giannis Spiliopoulos, Alexandros Troupiotis-Kapeliaris, Kostas Patroumpas, Nikolaos Liapis, Dimitrios Skoutas, Dimitris Zissis, Nikos Bikakis},
     pdfsubject={Trajectory Reconstruction, Maritime Traffic Analysis, Mobility Analytics, Trajectory Imputation, Vessel Monitoring, Spatial Data Management, Spatial Indexing},
     pdfkeywords={Trajectory Reconstruction, Mobility Analytics, Maritime Monitoring, Maritime Traffic Analysis, AIS Data, Spatial Data Management, Spatial Indexing}}

\begin{document}


 \title{Data-Driven Trajectory Imputation for Vessel Mobility Analysis}


 \author{Giannis	Spiliopoulos}
 \orcid{0000-0002-5003-7923}
  \affiliation{%
   \institution{University of the Aegean \\ \& Archimedes/Athena RC}
 \country{Greece}
 }

 \author{Alexandros \\ Troupiotis-Kapeliaris}
 \orcid{0000-0001-8726-6693}
\affiliation{%
   \institution{University of the Aegean \\ \& Archimedes/Athena RC}
 \country{Greece}
 }

 \author{Kostas Patroumpas}
 \orcid{0000-0003-3334-8671}
  \affiliation{%
   \institution{Archimedes/Athena RC}
  \country{Greece}
 }

 \author{Nikolaos Liapis}
 \affiliation{%
  \orcid{0000-0002-6118-5227}
   \institution{Hellenic Institute of Marine Technology}
 \country{Greece}
 }
 \author{Dimitrios Skoutas}
  \orcid{0000-0002-6118-5227}
 \affiliation{%
  \institution{Athena RC}
 \country{Greece}
 }

 \author{Dimitris Zissis}
 \orcid{0000-0003-2870-2656}
 \affiliation{%
   \institution{University of the Aegean}
 \country{Greece}
 }

 \author{Nikos Bikakis}
 \orcid{0000-0001-6859-1941}
 \affiliation{%
 \institution{Hellenic Mediterranean University \\ \& Archimedes/Athena RC}
 \country{Greece}
 }

 \renewcommand{\shortauthors}{}
 \renewcommand{\shorttitle}{}

 \begin{abstract}
     \input{TEX/abstract}

 \end{abstract}

 \keywords{Trajectory Imputation, Trajectory Reconstruction, Mobility Analytics, Maritime Monitoring, Maritime Traffic Analysis, AIS Data}

 \maketitle

\input{TEX/introduction}
 \input{TEX/related}

 \input{TEX/pol}

 \input{TEX/evaluation}

 \input{TEX/conclusion}

 \vspace{0.3cm}
 \noindent{\bf Acknowledgments.}
 This work has been partially supported by project MIS 5154714 of the National Recovery and Resilience Plan Greece 2.0 funded by the European Union under the NextGenerationEU Program.

 \bibliographystyle{ACM-Reference-Format}
 \bibliography{references}

 \end{document}

%% file: edbt-macros-v29-n3.tex

\def\EDBTISSN{2367-2005}
\def\EDBTISBN{978-3-98318-104-9}
\setcopyright{none}
\copyrightyear{\copyright\ 2026 Copyright held by the owner/author(s).
  Published on \url{OpenProceedings.org} under ISBN \EDBTISBN, series ISSN \EDBTISSN.
  Distribution of this paper is permitted under the terms of the
  Creative Commons license CC-by-nc-nd 4.0}
\acmYear{2026}
\acmDOI{}
\acmISBN{}

\acmConference[EDBT '26]{Extending Database Technology}{24-27 March 2026}{Tampere (Finland)}

\settopmatter{printacmref=false, printccs=false, printfolios=false}

\newsavebox{\ximagebox}
\newlength{\ximageheight}
\newsavebox{\xglyphbox}
\newlength{\xglyphheight}
\newcommand{\xbox}[1]%
  {\savebox{\ximagebox}{#1}%
  \settoheight{\ximageheight}{\usebox{\ximagebox}}%
  \savebox{\xglyphbox}{\color{white}\char32}%
  \settoheight{\xglyphheight}{\usebox{\xglyphbox}}%
  \raisebox{\ximageheight}[0pt][0pt]{\raisebox{-\xglyphheight}[0pt][0pt]{%
    \makebox[0pt][l]{\usebox{\xglyphbox}}}}%
    \usebox{\ximagebox}%
    \raisebox{0pt}[0pt][0pt]{\makebox[0pt][r]{\usebox{\xglyphbox}}}}


\newsavebox{\LogoBox}
\sbox{\LogoBox}{\includegraphics[height=1cm]{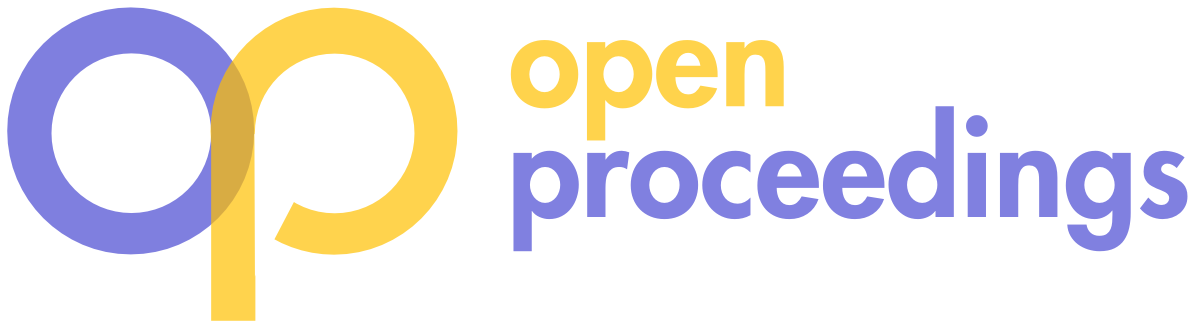}}

\fancypagestyle{firstpagestyle}{%
  \fancyhf{}%
  \fancyfoot{}%
  \fancyhead[L,C]{}
  \fancyhead[R]{\href{https://OpenProceedings.org/}{\raisebox{15pt}[0pt][0pt]{\xbox{\usebox{\LogoBox}}}}}
  }

\hypersetup{%
  pdfcopyright={CC-by-nc-nd},
  pdfpublisher={OpenProceedings.org -- University of Konstanz, Germany},
  pdfcreator={LaTeX with acmart, hyperref, and EDBT modifications},
  pdfvolumenum={29},
  pdfissuenum={2},
  pdfisbn={\EDBTISBN},
  pdfissn={\EDBTISSN}
}

%% file: TEX/abstract.tex
Modeling vessel activity at sea is critical for a wide range of applications, including route planning, transportation logistics, maritime safety, and environmental monitoring. 
%
Over the past two decades, the Automatic Identification System (AIS) has enabled real-time monitoring of hundreds of thousands of vessels, generating huge amounts of data daily. One major challenge in using AIS data is the presence of large gaps in vessel trajectories, often caused by coverage limitations or intentional transmission interruptions. These gaps can significantly degrade data quality, resulting in inaccurate or incomplete analysis.
State-of-the-art imputation approaches have mainly been devised to tackle gaps in vehicle trajectories, even when the underlying road network is not considered. But the motion patterns of sailing vessels differ substantially, e.g., smooth turns, maneuvering near ports, or navigating in adverse weather conditions. In this application paper, we propose HABIT, a lightweight, configurable \underline{H}3 \underline{A}ggregation-\underline{B}ased \underline{I}mputation framework for vessel \underline{T}rajectories. This data-driven framework provides a valuable means to impute missing trajectory segments by extracting, analyzing, and indexing motion patterns from historical AIS data.
Our empirical study over AIS data across various timeframes, densities, and vessel types reveals that HABIT produces maritime trajectory imputations performing comparably to baseline methods in terms of accuracy, while performing better in terms of latency while accounting for vessel characteristics and their motion patterns.

%% file: TEX/introduction.tex
\section{Introduction}
\label{sec:introduction}


More than 80\% of the world trade volume is carried by sea and is expected to grow on average 2\% annually by 2030\footnote{\href{https://unctad.org/publication/review-maritime-transport-2024}{https://unctad.org/publication/review-maritime-transport-2024}}, highlighting the impact of maritime operations on the world economy, environmental protection, and geopolitical stability.
The Automatic Identification System (AIS) currently installed on board 400K vessels worldwide offers a valuable means to monitor maritime activity. Capturing AIS signals relayed by vessels not only enables their real-time tracking but also safe navigation; but the vast volume of accumulated data also provides significant opportunities for a variety of operations, ranging from detailed traffic density maps and efficient port management to cost-effective route planning.

\stitle{Limitations of AIS Data.}
Yet, AIS data come with some inherent drawbacks, not only because of incomplete coverage, signal loss, and irregular reporting frequency of AIS messages, but sometimes even due to deliberate switch-off of tracking devices or intentional transmission of false information (spoofing) raising security concerns (e.g., collision) or suspicions (e.g., smuggling). Indeed, AIS positional messages may be recorded sparsely, resulting in \emph{gaps} with no information on the vessel's whereabouts for considerable time (even hours or days). Hence, reconstructing the complete trajectory of a vessel as a sequence of timestamped locations may not be possible, as its exact course may not always be known. The higher the sparsity of the sampled positions, the lower the accuracy and quality of the trajectory data. In particular, multiple such gaps may be observed, greatly diminishing the value of such data for reliable analysis. Certain applications (e.g., traffic monitoring, maritime surveillance) require high-quality trajectories, i.e., more `dense' trajectories without gaps for effective analyses.  


\stitle{Challenges in Maritime Domain.}
To alleviate this deficiency, \emph{trajectory imputation} techniques \cite{10.1145/3557915.3560942, 10.14778/3632093.3632105, 10.1145/3589132.3625620, 10.1145/3555041.3589733} may be applied to insert artificial points along these gaps in order to restore continuity of trajectory segments. The goal is to obtain imputed trajectories that resemble actual paths followed by vessels, and augment the quality of trajectories by eliminating discrepancies in their sampling rate.
Most existing methods for trajectory imputation ({\it aka} trajectory interpolation, trajectory completion, trajectory restoration, trajectory recovery \cite{10.1145/3656470}) rely on matching the trajectories on an underlying (typically: road) network, implicitly assuming that such a network is available and reliable. 
However, these methods are not particularly suitable for maritime trajectories due to several characteristics of vessel movement, such as:
{(1)}~Even though some types of vessels (e.g., passenger, tanker) follow strict navigation rules and recurring patterns, no fixed network can be safely assumed at sea;
{(2)}~The varying and occasionally low sampling rate of AIS positional messages makes it all the more difficult for such methods tailored for frequently reported GPS trajectories (typically from moving vehicles); and
{(3)}~Most importantly, the way vehicles move differs substantially from the motion patterns of vessels at sea, characterized by smooth turns especially for large ships, complex maneuvering near ports or at anchorage points, ad-hoc navigation during adverse weather conditions, and more.

\begin{figure}
    \centering
    \includegraphics[width=0.95\linewidth]{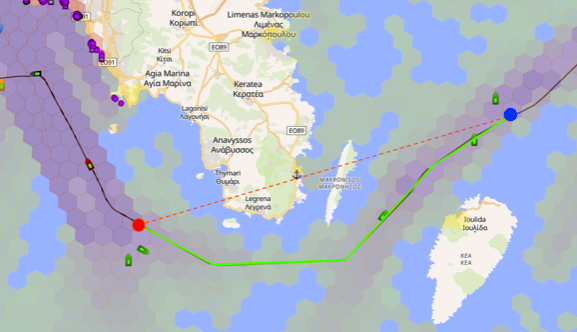}
    \caption{Imputing a gap along a vessel trajectory.}
    \label{fig:example}
\end{figure}

\stitle{Applications.}
Imputed trajectories can serve as a valuable resource for critical operations for a wide spectrum of use cases, including:

\emph{Maritime Decision-Making.}
Imputed trajectories can support more effective maritime monitoring in several scenarios, such as enabling authorities to \textit{prioritize actions in congested or environmentally sensitive areas}, generating path alternatives that consider weather condition predictions \cite{troupiotis2025dynamic}
or coordinating emergency response during search and rescue operations \cite{DBLP:conf/edbt/0002ZST18}. 

\emph{Visual Analysis of Vessel Traffic.} By removing gaps and noise from AIS data, it becomes possible to generate, update, and maintain more accurate \textit{density maps}
. Such maps (Fig.~\ref{fig:example}) provide an effective means of deriving rich insights into vessel traffic behavior and its evolution, offering an intuitive visual representation capable of encapsulating valuable information and supporting identification of patterns, anomalies, and other relevant phenomena \cite{rs15215080}.

\emph{Anomalous or Suspicious Maritime Behavior.}
Analyzing patterns from complete trajectories may reveal anomalous or suspicious behavior that indicates \textit{fraud or criminal activity, like package picking or vessel rendezvous} \cite{10.1007/s10707-016-0266-x}.


\stitle{Our Approach.} We introduce HABIT\footnote{\label{foot:github}\href{https://github.com/M3-Archimedes/HABIT}{https://github.com/M3-Archimedes/HABIT}}, an \underline{H}3 \underline{A}ggregation-\underline{B}ased \underline{I}mputation framework specifically for vessel \underline{T}rajectories.
This is an efficient, data-driven and scalable  approach for gap imputation in maritime trajectories relying on spatial aggregates of AIS positional reports computed over H3\footnote{\href{https://h3geo.org}{https://h3geo.org}} hexagon cells. 
State-of-the-art imputation approaches have been devised mainly to tackle gaps across vehicle trajectories even if the underlying road network may not be taken into account. Emanating from the recently proposed \emph{Patterns of Life} framework \cite{spiliopoulos2024patterns}, we suggest that extracting vessel motion patterns over large historical AIS data can offer improved opportunities for imputing the missing trajectory segments. 
Our data-driven, grid-based method can identify the underlying vessel movement relationships and  mobility patterns, and also keep per cell statistics, like distinct counts of vessels, number and distance of transitions between cells, etc.
Once such features are indexed, they can be queried in order to suggest possible  paths that impute the observed gaps.

HABIT focuses on filling gaps caused by AIS coverage issues rather than by deliberate AIS deactivation.
Gaps due to insufficient coverage affect most vessels in a given time and area, usually for a short duration, with vessels continuing their usual behavior. In contrast, intentional deactivation blocks vessel identification, location, and activities through tracking systems. Such cases deviate from typical behavior and are therefore less relevant to historical data in an area.     
We assume that statistics from recent historical AIS data, calculated over regular time intervals for a region, are representative of typical maritime traffic in that area. 
Consequently, any missing segments of historical trajectories can be reconstructed using such statistics.

 
It is important that the imputed paths must be \emph{navigable}, i.e., they do not cross coastlines, islands, protected zones, or other barriers and do not include frequent abrupt turns. Thus, the suggested route can be followed in practice by vessels, as denoted with a green line for the gap between the blue and the red point in Figure~\ref{fig:example}; in contrast, the straight path obtained from linear interpolation is clearly not navigable. The type of the vessel can be taken into account, since those of large size or draught (e.g., passenger or cargo ships) may not navigate through narrow straits or shallow waters. Our empirical study on real-world AIS data of varying timespan, spatial coverage, and vessel types confirms that imputing maritime trajectories must take into account the unique characteristics of vessels and their motion patterns leading to considerable improvements in terms of accuracy, performance, and scalability. HABIT provides a practical advantage, as it is based on lightweight libraries and frameworks, requiring significantly less computational power for both the framework construction and inference. Additionally, the inclusion of data-driven corrections of the paths allows for more realistic and grounded trajectory reconstructions.






\stitle{Outline.} The remainder of this paper proceeds as follows. Section~\ref{sec:related} offers necessary background on vessel monitoring and surveys related work on trajectory imputation. In Section~\ref{sec:modeling} we introduce HABIT, our framework for capturing vessel traffic from historical AIS data. Results from our evaluation study are reported in Section~\ref{sec:evaluation}. Finally, Section~\ref{sec:conclusion} concludes the paper with pointers for future work.

%% file: TEX/related.tex
\section{Background \& Related Work}
\label{sec:related}

\stitle{Vessel Monitoring.} 
Tracking the self-reported positions of sailing vessels through the Automatic Identification System (AIS) allows for advanced studies on human activity at sea. AIS messages relayed from vessels include their unique Maritime Mobile Service Identity ($MMSI$) number, the current coordinates ($LON$gitude, $LAT$itude, Speed Over Ground ($SOG$), Course Over Ground ($COG$), etc. Crucially, the timestamp is assigned when the message is received. The large data volumes collected daily require suitable tools for processing and modeling to provide a compact description of vessel traffic \cite{spiliopoulos2024patterns}. Moreover, specific mechanisms should be developed to ensure the quality of the resulting data, as AIS come with some inherent drawbacks (mainly due to coverage and signal loss issues). 

AIS centralized systems typically rely on a combination of  terrestrial and satellite radio-frequency receivers to acquire data \cite{yang2019big}. Coverage of terrestrial receivers is governed by radio frequency transmissions specifications, typically expanding for up to 20 nautical miles and is limited by the ground topology and weather conditions. On the other hand, satellite receivers, even though they have a much larger reception footprint, suffer from coverage issues mostly due to satellite presence, i.e., objects in orbit, and packet collision  on crowded areas. Given that AIS is a self-reporting system, any type of receivers can be subject to missing or false data from intentional mis-use/malicious use of AIS on-board. A recent analysis \cite{rs15215080} that compared AIS with Earth Observation detection showed that even in industry-level, centralized AIS databases, a significant amount of AIS messages may be missing in certain areas, indicating large gaps along the trajectories of vessels passing by. 

\stitle{Trajectory Imputation.} Gaps in reported locations are not observed only in vessel trajectories, but concern moving objects in general (vehicles, commodities, animals, etc.). Hence, state-of-the-art imputation methods attempt to insert artificial points between the real ones in order to suitably bridge such gaps. \emph{Network-based} methods like \cite{chen2023rntrajrec} leverage the structure of the underlying network (usually, roads) to infer potential paths for the missing segments. 

In contrast, \emph{network-less} imputation techniques are based on the positional data. Next, we first outline such data-driven methods specifically addressing gaps in GPS trajectories in urban settings. TrImpute \cite{10.1145/3557915.3560942} assumes no knowledge of the underlying road network.  For each trajectory segment to be imputed, it first collects candidate points from existing trajectories nearby, and then picks those that form the shortest path between the end points of the segment. 
In the same spirit, GTI \cite{10.1145/3589132.3625620} proposed an improved network-agnostic, but computationally demanding method that exploits existing knowledge of neighboring trajectories to densify a target trajectory. It creates a connected directed graph from the sparse trajectories to enable finding the shortest path using Dijkstra between every two successive locations
by traversing the graph nodes. 

ML models have also been applied for trajectory imputation, but incur high computational complexity. For instance, TERI \cite{10.14778/3632093.3632105} proposes a framework for trajectory recovery without any prior information about the recovery positions. Assuming irregular time intervals between such positions, it first detects when they occurred, and then proceeds to the imputation of the missing data points. In each stage, it deploys a model (RETE) based on Transformer encoder architecture with a learnable Fourier encoding module to better model spatial and temporal correlations, and integrates collective transition pattern learning and trajectory contrastive learning to effectively capture sequential transition patterns. Besides, Kamel \cite{10.1145/3555041.3589733} introduced a scalable trajectory imputation framework that inserts additional realistic trajectory points to improve accuracy. It maps the trajectory imputation problem to finding the missing word problem in NLP, thus employing the widely used BERT model with several important enhancements. It adds spatial-awareness to BERT operations, adjusts trajectory data to be closer to the nature of language data, and offers multipoint imputation ability. PaLMTO~\cite{10591657} explores the use of probabilistic language models based on $N$-grams for trajectory imputation, i.e., predict the next token (i.e., point) in the trajectory by considering the previous $N-1$ tokens as context. It employs a grid-based spatial representation oblivious of any underlying road network, and converts trajectory points into tokens corresponding to the grid cell where they appear to train models of different sizes. 


However, all aforementioned techniques were designed for and tested against GPS trajectories in urban settings. As we stressed above, they may not produce reliable paths against AIS data, thus imputation methods have been proposed specifically for vessel trajectories. DAISTIN~\cite{10.1145/3609956.3609961} constructs a graph from the raw AIS data to represent movement specifically from oil tankers. To fill in a gap between two locations along a given trajectory, it returns the shortest path. However, it has not been tested on other vessel types. 
In \cite{10.1145/3637528.3672086}, a physics-guided diffusion probabilistic model was suggested for long-term  gaps (up to 10 hours) regarding tanker and cargo vessels. It reconstructs trajectories from white noise using conditional information, and also specifies kinematic constraints to enhance trajectory smoothness when the diffusion model is trained. Similarly, TrajDiff~\cite{WANG2025107279} employs diffusion models by adding noise to generate latent AIS data representations and thus enhance learning of trajectory traffic distribution and vessel motion patterns. The denoising process employs known trajectory points and motion features as conditions in order to generate more realistic trajectory segments along long-lasting gaps  (e.g., 10 hours). 
Most recently, MH-GIN~\cite{10.14778/3773749.3773756} constructs a multi-scale graph-based network to explicitly model dependencies between heterogeneous AIS attributes (not just locations) for more accurate imputation of missing values through graph propagation. 
In contrast, some works address the trajectory imputation problem by introducing additional data sources. For example, \cite{WuTTZZS25} employs a data fusion approach to incorporate coastal camera detections, resulting in more complete trajectories.

In this work we build upon concepts presented in \cite{spiliopoulos2024patterns}, employing an H3 spatial index for information compression and construction of a transition graph. Compared to its distributed  processing architecture on Apache Spark, we propose a simpler technological stack that can leverage DuckDB\footnote{\url{https://duckdb.org}}, a modern analytics database, to efficiently process data in-memory. Most importantly, we specifically examine the application of those concepts to the trajectory imputation problem.
We introduce a data-driven mechanism to mitigate information loss from the use of the H3 grid and analyze and compare our method against the original path and state-of-the-art methods across various configurations and datasets.

%% file: TEX/pol.tex
\section{HABIT: H3 Aggregation-based Imputation for Vessel Trajectories}
\label{sec:modeling}





In this section we present HABIT
(\underline{H}3 \underline{A}ggregation-\underline{B}ased \underline{I}mputation framework for vessel \underline{T}rajectories)$^{\ref{foot:github}}$, which is built on top of open-source tools such as DuckDB and NetworkX.
HABIT is a data-driven and scalable approach for gap imputation in maritime trajectories, relying on spatial aggregates of AIS positional reports computed over H3 hexagon cells.



Overall, our framework is comprised of four main phases:~
\begin{enumerate}
\item
\textit{Data preprocessing \& trip segmentation} (Section~\ref{sec:prepr}): Erroneous data is filtered out and raw positional reports are cleaned, analyzed and organized by trip.

\item
\textit{Graph generation} (Section~\ref{sec:graphgen}):
Position embeddings are computed using the spatial H3 index.
The transitions between cells are first organized into  an edge list and then assembled into a graph.

\item
\textit{Trajectory imputation} (Section~\ref{sec:imput}):
A shortest-path algorithm is used to identify the points of the missing trajectory segments, leveraging graph statistics.
Additionally, a data-driven method is applied to more accurately project between H3 cells and coordinates.


\item
\textit{Trajectory simplification} (Section~\ref{sec:reconstruct}):
A smoothing process is applied to the imputed trajectory to generate a realistic and navigable path.

\end{enumerate}



\subsection{Data Preprocessing \& Trip Segmentation}
\label{sec:prepr}

In the initial phase, we segment the entire trajectory of each vessel into separate trips and perform a data cleaning process.
In order for the trips to be identified, the AIS raw data is analyzed by considering several factors such as time, location, speed changes, and heading. This analysis allows us not only to identify individual trips but also to detect various motion events, which are semantically annotated.

In our approach we consider as a \emph{trip} the subsequence of AIS locations between two successive stops or gaps:
\begin{itemize}
\item A \emph{stop} indicates that the vessel remains stationary, i.e., its speed is $<0.5$ knots over a period of time. The starting location of such a stop (most typically in a port, but sometimes also at sea) marks the end of the current trip. Conversely, the last location in a stop signifies that the vessel has just departed on a new trip.

\item A \emph{communication gap} occurs when no AIS location has been received recently,
e.g., in the past $\Delta T = 30$ minutes. If such a gap occurs, the current trip is abruptly ended and a new trip will be assigned once communication with the vessel is resumed. Note that gaps lasting less than $\Delta T$ may still occur in the extracted trips.
\end{itemize}

To identify such stops and gaps, we employ a \emph{trajectory compression} framework~\cite{10.1007/s10707-022-00475-0} tailored for annotating AIS data\footnote{\href{https://github.com/M3-Archimedes/AIS-trajectory-annotation}{https://github.com/M3-Archimedes/AIS-trajectory-annotation}}.
By examining how the motion pattern (speed, heading) of a given vessel changes across time, this method can incrementally annotate selected positions that signify several types of mobility events (not only stops and gaps, but also turning points, slow motion, and speed change).
This process also applies filters that  eliminate much of the \emph{noise} inherent in the AIS messages, such as duplicate positions, invalid coordinates, delayed messages that distort the sequence, etc.
Once these annotations are obtained, noisy locations are discarded, and all remaining locations of a vessel between two successive stops or gaps are grouped together as a trip and its unique identifier \emph{TRIP\_ID} is assigned to each location.

Finally, it may occur that a trip designated by the aforementioned criteria falls within just one or at most two adjacent H3 cells at a given resolution. As these records primarily concern minor, non-essential local displacements (e.g., sea drift), they are excluded from further consideration in our analysis.



\subsection{Graph Generation}
\label{sec:graphgen}


Unlike land transportation networks, which can be considered static over short periods, sea transportation networks are dynamic. Even routes between ports can swiftly change due to weather, traffic and geopolitical factors. To capture the collective behavior of vessels, raw AIS data is projected onto an H3 grid and local transitions are grouped among grid cells. This process constructs a data-driven maritime network from AIS data in the form of a graph, which can then be utilized to take advantage of existing graph-based path algorithms.





\stitle{Data processing \& Indexing.}
AIS messages are processed as follows using a common table expression (CTE) in DuckDB:

\begin{enumerate}
    \item
Trip data obtained from the initial phase is read from the source files, and the H3 spatial resolution is defined. 

\item
The messages are organized into distinct groups based on the trip identifier (TRIP\_ID).

\item
Each message is  assigned to its corresponding H3 cell ($cl$) at the specified resolution and augmented with its preceding H3 cell ($lag\_cl$) along the trip sequence.

 \item
The data is grouped twice, once per H3 cell ($cl$) and once by the ($lag\_cl$,$cl$) tuples. In both cases, aggregated statistics are computed for each group.
\end{enumerate}

A challenging and critical issue in  indexing step is to determine the resolution ($r$) of the H3  grid. While some works, particularly in urban settings, may use resolutions of just a few meters \cite{10.1145/3555041.3589733}, AIS trajectories typically span much larger distances. If the grid is too fine, it becomes difficult to extract reliable descriptive statistics, as each cell would contain only a small number of AIS points. Conversely, choosing a larger cell size reduces detail, causing the resulting paths to align more with the grid layout than with the actual vessel tracks.

\stitle{Statistics Computations.}
Statistics are calculated using \textit{DuckDB built-in functions}
to compute exact and approximate metrics.

For each H3 cell group $cl$ we compute:
\begin{itemize}
    \item $count(*)$:  the total number of AIS records within cell $\text{\textit{cl}}$.

\item $apprx\_count\_distinct(VESSEL\_ID)$: the number of distinct vessels passing through  cell $\text{\textit{cl}}$.

\item $median(lon)$ and $median(lat)$: the median longitude and latitude, respectively.

\item $median(sog)$ and $median(sog)$:
the median speed and course over ground.
\end{itemize}





Then, for each H3 cell tuple group ($lag\_cl$,$cl$), we compute:
\begin{itemize}
\item  $apprx\_count\_distinct( TRIP\_ID)$: the number of transitions from cell $\text{\textit{lag\_cl}}$~to~$\text{\textit{cl}}$.

\item ~$h3\_grid\_distance(lag\_cl, cl)$:  the distance of transition expressed in H3 cells.

\end{itemize}

%
%
%








\stitle{Graph Construction.}
The H3 index along with the computed statistics are used to construct a weighted graph\footnote{The graph is constructed using the Python library \mbox{\url{https://networkx.org}}.}:

\begin{itemize}
\item The \emph{nodes} of the graph are identified by the corresponding H3 cells present in the edge list and are associated with the following attributes, as computed in the previous step:
        (a)~Median Longitude;
        (b)~Median Latitude;
        (c)~Total number of messages; and
        (d)~Number of distinct vessels. 

\item The \emph{edges} that connect two nodes based on their H3 index identifiers (typically $cl$ and $lag\_cl$), are defined by the transitions found in the trajectories, with the condition that $lag\_cl\neq cl$. The respective edge weight corresponds to the aggregated number of transitions.

\end{itemize}

\subsection{Trajectory Imputation}
\label{sec:imput}


Initially, we project the coordinates of the gap start and end points onto their corresponding cells in the H3 grid.
In order to be able to exploit the underlying graph topology, both the start and end H3 cells must correspond to graph nodes.
If the cells do not correspond to graph nodes, a nearest-neighbor search is performed to find the closest cell that does.
As a result, the start and end points of a gap are mapped to two graph nodes.

Subsequently, an A* search identifies the shortest path by minimizing the number of transitions, which effectively reveals the most frequent path according to historical data.
At this stage, the resulting imputed path comprises the most frequent (in terms of transitions) sequence of nodes (H3 cells) for the gap.
To provide a solution in real-world coordinates, we perform the inverse projection of the H3 imputed path back to longitude and latitude tuples.
While the most straightforward approach would use the geometric center of each cell as its coordinate representation, it can easily result in positions that fall outside the vessels' common movement areas or even on land regions.
To address this, we employ a data-driven approach, computing the median coordinates for each cell during graph construction, according to the AIS positions.

\begin{figure}
    \centering
    \includegraphics[width=0.85\linewidth]{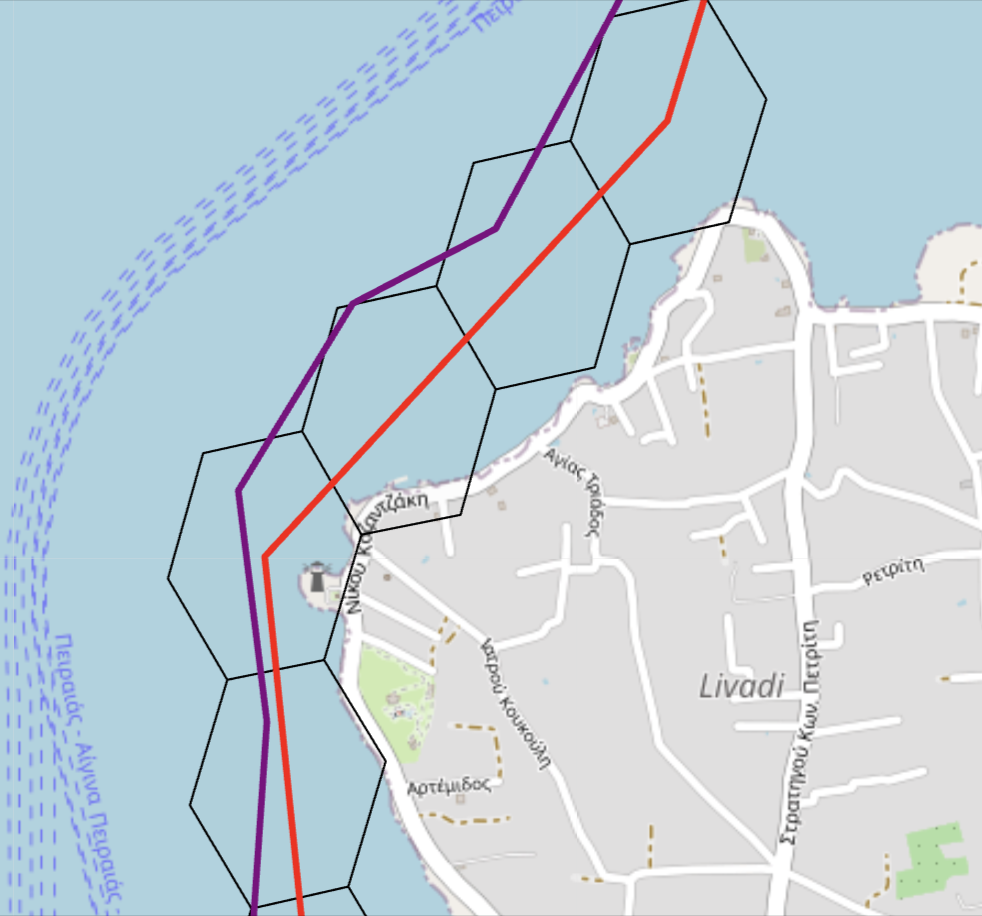}
    \caption{Inverse projection of the imputed path cell sequence back to real-world coordinates using (a) cell centers [red]; and (b) a data-driven (median) method [purple].}
    \label{fig:weightedvspol}
\end{figure}

Utilizing the median over the average coordinates ensures the derived position is grounded in a historically valid vessel location, thus aligning with true movement patterns (Figure~\ref{fig:weightedvspol}).
HABIT can support either projection option for the cells, depending on the chosen configuration parameter $p$.
We denote the geometric center as $c$ and
the median as~$w$.


\subsection{Trajectory Simplification}
\label{sec:reconstruct}

The final stage includes a simplification of the imputed trajectory, since producing realistic and navigable paths is paramount in maritime navigation.
This requires paths with infrequent significant turns, that can be defined by a series of straight-line segments between waypoints.
This need for smoothness, which is distinct from urban mobility, originates from the limited maneuverability of most vessels, especially large ones. Since paths derived from a grid-based approach inherently follow the structure of the underlying grid, they often introduce multiple sudden `zigzags' even if they accurately track the sequence of cells visited.
For this process   the Ramer-Douglas-Peucker (RDP) algorithm \cite{visvalingam1990douglas} is utilized as a highly effective trajectory simplification and generalization technique.
The selected \emph{tolerance} $t$ is a crucial parameter governing RDP's operation, as it dictates the maximum allowable deviation from the given path. Larger thresholds yield smoother paths but may diverge to the point of violating the graph schema or crossing over land areas. Conversely, choosing a smaller tolerance may fail to achieve acceptable path compression and could retain undesirable abrupt turns between neighboring cells.

%% file: TEX/evaluation.tex
\section{Experimental Analysis}
\label{sec:evaluation}

\subsection{Setup}
To evaluate our framework, we use real vessel tracks from AIS in two distinct European regions.
The evaluation process was divided into a two-step approach:
(a) fine-tuning the proposed method's parameters by examining their impact on the reconstructed trajectories, and
(b) comparing the selected configurations against  baseline methods.
For the latter analysis, we focused on both the accuracy of the generated imputations and the overall performance in terms of query latency and memory requirements. Finally, we include an analysis for HABIT's accuracy on larger gap durations.

\stitle{Datasets.}
We experimented with AIS data from two data sources, which cover different configurations and scenarios.
\textit{Danish Maritime Authority AIS}\footnote{\url{https://www.dma.dk/safety-at-sea/navigational-information/ais-data}}: We extracted all raw AIS positions regarding \emph{passenger} ships for a 3-month period (Jan--March 2024) around Denmark. Two datasets were subsequently derived:
\textbf{DAN}: an extended set of selected trips between 10 specific ports of the region (Denmark, Sweden, Germany). This scenario concerns selected routes for a specific vessel type across a broader area.
\textbf{KIEL}: all trips between the ports of Kiel (Germany) and Gothenburg (Sweden). This scenario concerns itineraries following a specific route.

 \textit{AegeaNET AIS}\footnote{\url{https://smartmove.aegean.gr/aegeanet}} (\textbf{SAR}): We created a full month dataset  from the broader area of Saronic gulf, near the port or Piraeus (Greece), including the tracks of all vessel types. This scenario concerns all traffic in a maritime area with varying quality of AIS reception, without any filtering of vessel types or routes.
Details of these datasets are listed in Table~\ref{tab:data}.




\begin{table}[!t]
\centering
\caption{Characteristics of the AIS datasets.}
\label{tab:data}
\vspace{-6pt}

\sisetup{group-separator={,}}
\setlength{\tabcolsep}{4pt}
\begin{tabular}{ll
                S[table-format=4]
                S[table-format=7]
                S[table-format=5]
                S[table-format=4]}
\toprule
\textbf{Dataset} & \textbf{Type} &
\textbf{Size (MB)} &
\textbf{Positions} &
\textbf{Trips} &
\textbf{Ships} \\
\midrule
DAN  &  {Passenger} & 786 & 4384003 & 1292  & 16 \\
KIEL &  {Passenger}                           & 145 & 806498  & 86    & 2  \\
SAR  & All                         & 141 & 1171162 & 20778 & 2579 \\
\bottomrule
\end{tabular}
\end{table}

To assess the imputation results, we introduced \emph{synthetic gaps} of fixed duration: 60, 120, and 240 minutes (default: 60 minutes).
A single gap was placed randomly within each trip. The original trips (without artificial gaps) serve as ground-truth. In evaluating our approach, 70\% of the trips were utilized to construct the underlying graph structures required for subsequent inference.
The remaining 30\% were used for accuracy and performance testing.



\stitle{Competitors.} We compare our H3-based approach HABIT with GTI~\cite{10.1145/3589132.3625620}, a highly accurate, state-of-the-art imputation method. This method relies on two distance parameters during graph creation: $rm$ (radius in meters) and $rd$ (radius in degrees), which filter candidate edges between points. In our experimentation we used values relative to the vessel size to reflect typical spacing in vessel traffic.
Initially, we had also included  PaLMTO~\cite{10591657} in our experiments, as it yielded models comparable in size to the most refined HABIT configuration. However, during inference, the resulting models failed to provide sufficient imputation results, frequently exceeding the time limit and falling into a timeout; thus, results from PaLMTO are not reported next.
Finally, we include the simple straight-line interpolation (SLI) as a baseline, which naively connects the two endpoints of a gap with a direct line segment.



\begin{figure}[t]
    \centering
    \includegraphics[width=0.9\linewidth]{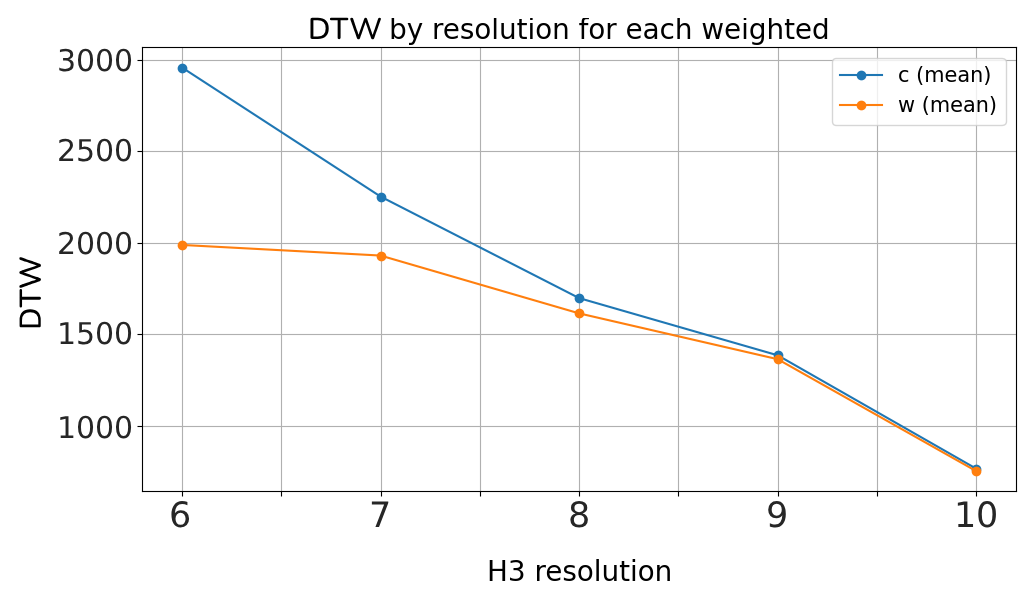}

    \caption{HABIT  accuracy (DTW) at different H3 resolutions and $p$ options (H3 cell center / Data median) [DAN Dataset].}
    \label{fig:weighted}
\end{figure}

\stitle{Evaluation Metrics.}
We assess the methods in terms of  accuracy and performance.
Regarding \textit{accuracy}, we use the
\textit{Dynamic Time Warping} (DTW) \cite{berndt1994using} that indicates the average distances between the imputed and original paths. For meaningful DTW measurements, the imputed trajectories were interpolated, ensuring that consecutive positions were at most 250m apart.
Regarding \textit{performance}, we measure the average execution latency for imputation queries and the memory footprint of the constructed frameworks.
We conducted our experiments on a server equipped with a 13th Gen Intel Core i9-13900K  (32 cores), 128GB of RAM, running Ubuntu 22.04.5 LTS.



\subsection{Parameter Fine-tuning for HABIT}

The proposed H3-based method HABIT relies heavily on a set of parameters to impute the requested paths. These parameters define a refined level of spatial discretization and path simplicity, and establish an operationally feasible solution via lightweight optimizations.
Several configurations were tested on the {DAN} dataset, which contains long-distance trips between multiple pairs (origin-destination) of ports.


In determining the H3 grid resolution ($r$) for the graph construction (Section~\ref{sec:graphgen}), we should consider that the average length of commercial vessels can be up to 300 meters (or even more in some cases), which in their trips can cover hundreds of kilometers. Hence, we examined $r$ values from 6 up to 10. As expected, higher resolutions produce more faithful results (Figure~\ref{fig:weighted}), but at the same time require more memory space for maintaining the graphs (Table~\ref{tab:sizes}).

Furthermore, as graph imputation results (Section~\ref{sec:imput}) are obtained as H3 cells, we studied the inverse projection process, according to the projection configuration parameter ($p$).
As observed in Figure~\ref{fig:weighted}, selecting a median center can provide a more refined result, closer to the real paths the vessels actually followed. This is especially apparent in coarser resolutions, where the displacement inside the cells is considerably larger.




 \begin{table}[]
\centering
\caption{Framework storage size (MB).}
\label{tab:sizes}
\vspace{-6pt}

\sisetup{
  round-mode = places,
  round-precision = 2,
  table-number-alignment = center
}
\setlength{\tabcolsep}{8pt}

\begin{tabular}{ll
                S[table-format=4.2]
                S[table-format=4.2]}
\toprule
\textbf{Method} & \textbf{Configuration} & \textbf{KIEL} & \textbf{SAR} \\
\midrule
\multirow{5}{*}{\textbf{HABIT}}
  & $r=6$  & 0.06   & 0.22   \\
  & $r=7$  & 0.29   & 0.59   \\
  & $r=8$  & 1.54   & 2.96   \\
  & $r=9$  & 8.20   & 18.03  \\
  & $r=10$ & 37.28  & 57.40  \\
\midrule
\multirow{3}{*}{\textbf{GTI}}
  & $rd=10^{-4}$  & 50.24    & 115.19   \\
  & $rd=5 \!  \cdot \!  10^{-4}$ & 369.41   & 3541.89 \\
  & $rd=10^{-3}$  & 1428.77  & \underline{4844.12} \\
\bottomrule
\end{tabular}
\end{table}

\begin{table}[]
\centering
\caption{Effect of simplification on the imputed trajectories: count of positions  (\textit{cnt}), average (\textit{Avg rot}) and maximum rate of turn (\textit{Max~rot}) and number of turns exceeding 45$^o$. Measurements are averages over all paths, with varying  \textit{tolerance} $t$ at two distinct H3 \textit{resolutions}~$r$ [DAN dataset].}
\label{tab:simplify}
\vspace{-6pt}

\sisetup{
  round-mode          = places,
  round-precision     = 2,
  table-number-alignment = center
}
\setlength{\tabcolsep}{8pt}

\begin{tabular}{cc
                S[table-format=3.2]
                S[table-format=2.2]
                S[table-format=3.2]
                S[table-format=2.2]}
\toprule
\textbf{$r$} & \textbf{$t$} &
\textbf{cnt} &
\textbf{Avg rot} &
\textbf{Max rot} &
\textbf{$>45^{\circ}$} \\
\midrule
\multirow{5}{*}{9}
  & 0    & 96.35 & 30.79 & 112.71 & 34.13 \\
  & 100  & 51.76 & 54.92 & 112.31 & 33.78 \\
  & 250  & 35.32 & 57.61 & 109.96 & 23.75 \\
  & 500  & 14.57 & 44.89 & 84.03  & 6.11  \\
  & 1000 & 6.93  & 34.32 & 56.05  & 1.64  \\
\midrule
\multirow{5}{*}{10}
  & 0    & 198.31 & 30.64 & 119.07 & 62.37 \\
  & 100  & 71.96  & 48.53 & 116.93 & 35.26 \\
  & 250  & 21.03  & 33.85 & 77.01  & 4.43  \\
  & 500  & 8.62   & 24.70 & 43.31  & 0.60  \\
  & 1000 & 4.67   & 19.85 & 27.38  & 0.09  \\
\midrule
\multicolumn{2}{r}{\textbf{Original}} &
595.63 & 6.55 & 110.79 & 33.84 \\
\bottomrule
\end{tabular}
\end{table}

\begin{figure}[t]
    \centering
    \includegraphics[width=0.99\linewidth]{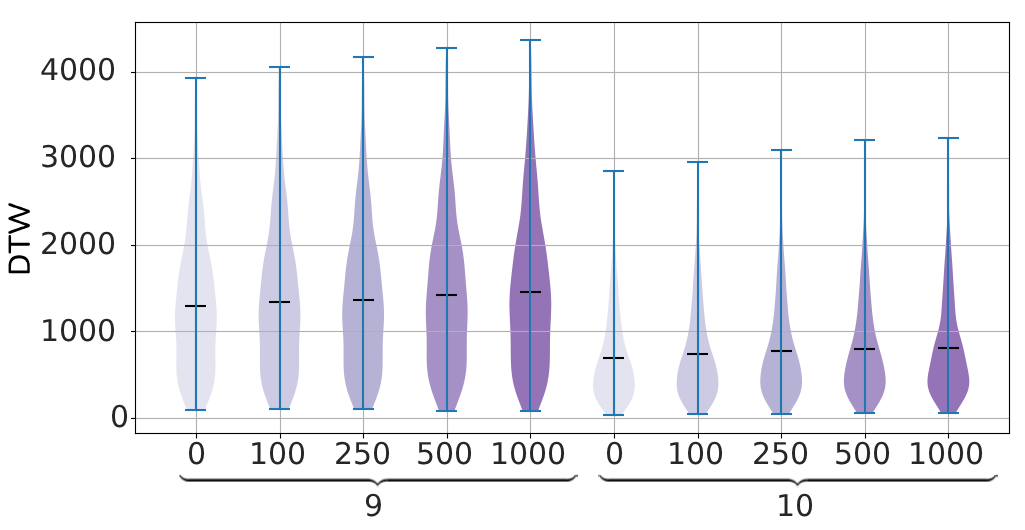}
    \caption{HABIT accuracy (DTW) for different simplification tolerances  ${t=\{0, 100, 250, 500, 1000\}}$, and   resolutions $r=\{9, 10\}$ [DAN Dataset].}
    \vspace{-6pt}
     \label{fig:simplify}
\end{figure}

Finally, as mentioned in Section~\ref{sec:reconstruct}, the RDP algorithm is employed to address this navigability of the imputed paths.
We investigated the effect of tolerance $t$ on the characteristics of the imputed trajectories, considering the number of positions per path, the average and maximum rate of turn (rot) and the number of turns exceeding 45$^o$.
Typically, we need to avoid very abrupt turns (over 45$^o$) while not oversimplifying the resulting paths.
From Table~\ref{tab:simplify} we concluded that a tolerance ranging from 100 to 250 meters can provide the optimal trade-off between path smoothness and compression. Besides, accuracy does not seem to be affected by the different $t$ values, as shown in Figure~\ref{fig:simplify}.
Nevertheless, accuracy does not appear to be affected by a finer resolution $r$ of the H3 grid.



\begin{figure}[h]
    \centering
    \includegraphics[width=0.99\linewidth]{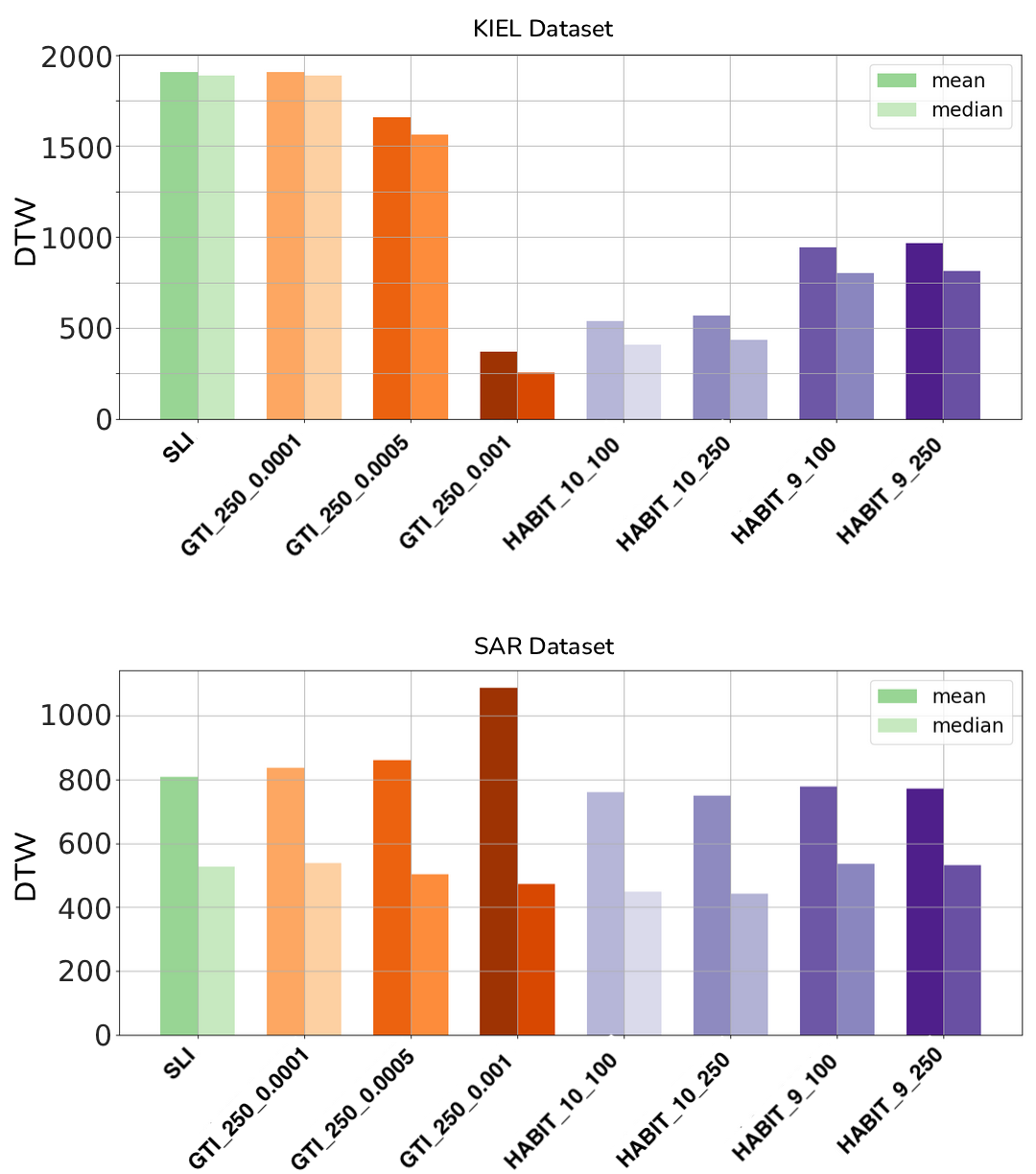}
    \caption{Sensitivity analysis on accuracy (mean \& median DTW) of the imputed paths with varying parameterizations for GTI ($rm$, $rd$) and HABIT ($r$, $t$) [KIEL \& SAR Dataset].
    }
    \label{fig:both_dtw}
\end{figure}

 \begin{figure*}[t]
    \centering
    \includegraphics[width=0.97\linewidth]{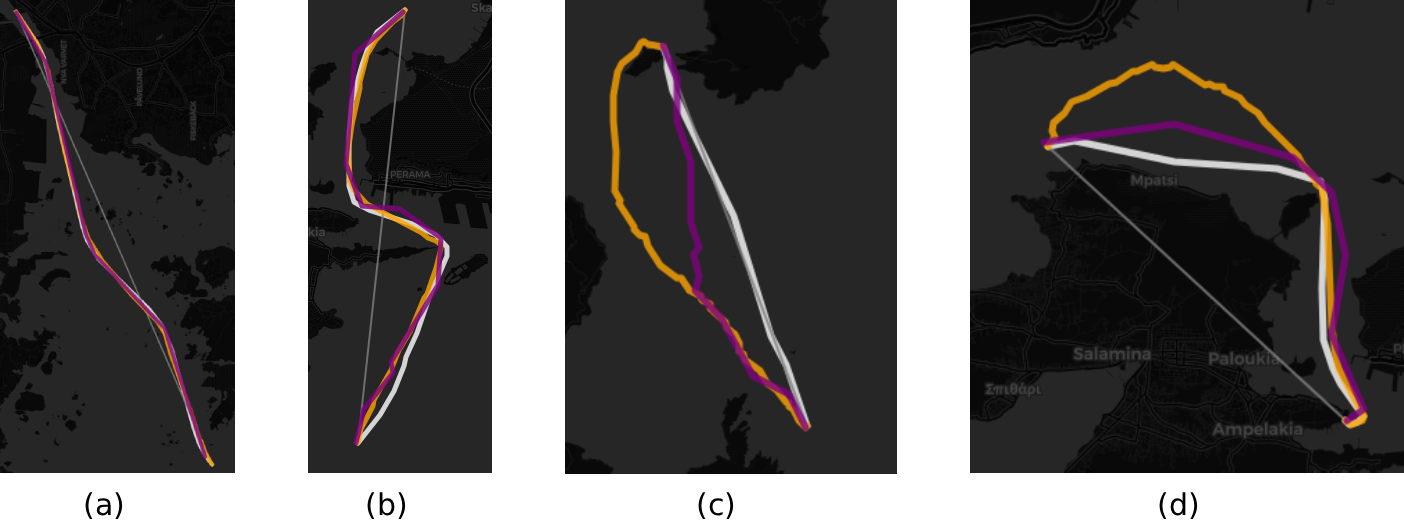}
    \vspace{-4pt}
    \caption{Indicative imputation results:
    original [white], HABIT  [purple], GTI [orange], and SLI [gray].}
    \label{fig:paths}
\end{figure*}

\subsection{Results}
Next, we report results from an empirical comparison of the trajectory imputation methods for a fixed gap of 60 minutes. Several configurations of both our proposed framework (HABIT) and the chosen state-of-the-art method (GTI) are included, along with the naive linear interpolation (SLI).
We selected the KIEL and SAR datasets to cover two distinct use case scenarios. The DAN dataset was excluded from the analysis because its size was prohibitive for generating the GTI models, and even downsized versions (by resampling locations per trip every 1 or 5 minutes) did not yield consistent results.

\subsubsection{Accuracy}

As depicted in Figure~\ref{fig:both_dtw}, the quality of the generated trajectories is very sensitive to parametrization.
Generally, both approaches are able to model local traffic patterns well, by limiting their search space using the raw AIS data during graph construction (Figure~\ref{fig:paths} a--b).

Trajectories in the KIEL dataset tend to follow a specific, confined route. Given this characteristic, both the GTI and HABIT methods achieve relatively low deviations from the original paths, particularly when contrasted with the SLI baseline, which cannot capture turning points. Since GTI is explicitly designed to follow real past tracks (which in this scenario all belong to the same route), it achieves a higher level of accuracy. On the other hand, HABIT is subject to inherent positional displacements within the cells, resulting in larger errors even in correct H3 sequence predictions.

With regard to the larger SAR dataset (covering all vessel types and all trips in the area), the two methods showcase different behaviors. This is due to the variety of alternative routes between two points, as found in the training data. More specifically, some configurations for GTI provide a comparable and in some instances slightly worse accuracy than SLI. For a more refined configuration ($rm=250m$, $rd= 0.001^o$), the memory requirements of GTI did not allow its computation. In its place, we included a different training configuration ($rm=500m$, $rd=0.001^o$, underlined in Table~\ref{tab:sizes}), which in turn did not perform as well, displaying larger errors.
Conversely, our proposed framework HABIT tends to remain relatively stable in terms of accuracy, avoiding extreme errors that would affect its mean score
(Figure~\ref{fig:both_dtw}). After inspecting the resulting paths, this phenomenon largely stems from GTI choosing outlier paths that do not move directly towards the destination point, especially when sailing outside of main shipping lanes (Figure~\ref{fig:paths} c--d). HABIT generally tends to follow the simpler path, minimizing the total weight but also following local patterns of movement, instead of single sampled trajectories; thus performing better in cases of typical sailing movement.

\begin{figure}
    \centering
    \includegraphics[width=0.99\linewidth]{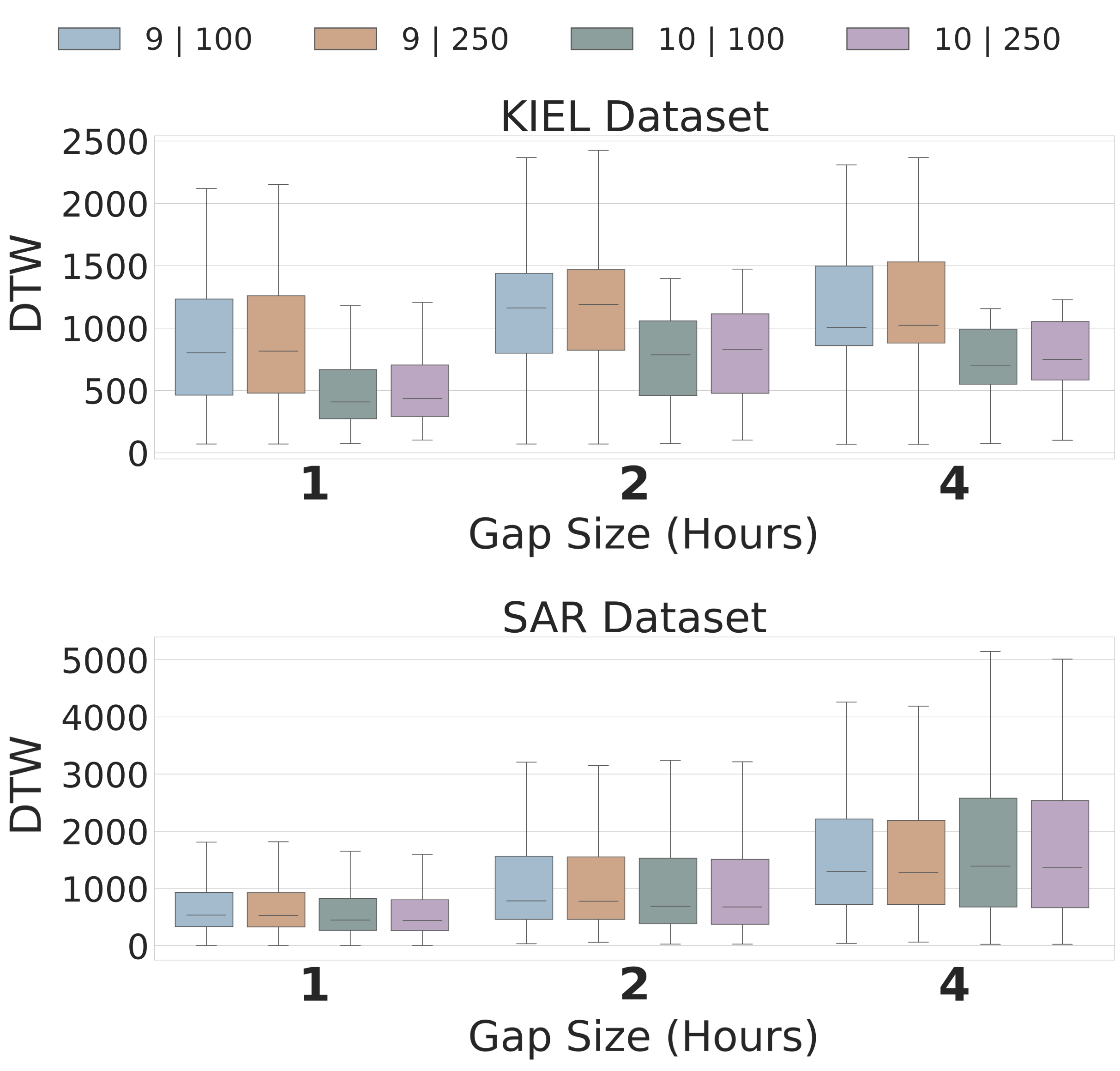}
    \caption{HABIT   accuracy (DTW) of imputed paths for gaps of 1, 2 and 4 hours duration, for different configurations relating to resolution and simplification factor ($r$|$t$) \newline [KIEL \& SAR Dataset].}
    \label{fig:extra}
\end{figure}

To further assess the applicability and efficiency of the proposed framework, we conducted additional experiments using HABIT's selected variations on larger gap durations (2 and 4 hours). As shown in Figure~\ref{fig:extra}, accuracy degrades with increasing gap size, as expected due to longer interpolation distances; however, the increase in median error is not proportional to the gap length, highlighting the robustness of the approach. This effect is mild in the KIEL dataset but more evident in SAR, where slightly higher medians are accompanied by pronounced outliers. These extreme errors are likely related to irregular trajectories of non-commercial or passenger vessels present in the SAR dataset. Importantly, the relative performance ranking of the variations remains generally consistent across gap sizes.

 \begin{table}[]
\centering
   \caption{Average and maximum query latency  (sec) for different configurations of HABIT ($r$, $t$) and GTI ($rm$, $rd$).}
\label{tab:latency}
\vspace{-6pt}

\sisetup{
  round-mode = places,
  round-precision = 3,
  table-number-alignment = center
}
\setlength{\tabcolsep}{3pt}

\begin{tabular}{lll
                S[table-format=1.3]
                S[table-format=1.3]}
\toprule
\textbf{Dataset} & \textbf{Method} & \textbf{Configuration} & \textbf{Avg} & \textbf{Max} \\
\midrule
\multirow{7}{*}{\parbox{2cm}{\centering \textbf{KIEL} \\ (26 gaps)}}
 & \multirow{4}{*}{\textbf{HABIT}}
 & $r=9,\; t=100$  & 0.024 & 0.041 \\
 &  & $r=9,\; t=250$  & 0.019 & 0.047 \\
 &  & $r=10,\; t=100$ & 0.071 & 0.121 \\
 &  & $r=10,\; t=250$ & 0.070 & 0.128 \\
\cmidrule(lr){2-5}
 & \multirow{3}{*}{\textbf{GTI}}
 & $rm=250,\; rd=10^{-4}$   & 0.261 & 0.281 \\
 &  & $rm=250,\; rd=5 \! \cdot \!  10^{-4}$ & 0.300 & 0.431 \\
 &  & $rm=250,\; rd=10^{-3}$   & 0.402 & 0.931 \\
\midrule
\multirow{7}{*}{\parbox{2cm}{\centering \textbf{SAR} \\ (1{,}871 gaps)}}
 & \multirow{4}{*}{\textbf{HABIT}}
 & $r=9,\; t=100$  & 0.032 & 0.202 \\
 &  & $r=9,\; t=250$  & 0.031 & 0.186 \\
 &  & $r=10,\; t=100$ & 0.139 & 0.963 \\
 &  & $r=10,\; t=250$ & 0.139 & 0.866 \\
\cmidrule(lr){2-5}
 & \multirow{3}{*}{\textbf{GTI}}
 & $rm=250,\; rd=10^{-4}$   & 0.492 & 0.550 \\
 &  & $rm=250,\; rd=5 \! \cdot \! 10^{-4}$ & 0.711 & 1.598 \\
 &  & $rm=500,\; rd=10^{-3}$   & 1.216 & 5.185 \\
\bottomrule
\end{tabular}
\end{table}

\subsubsection{Performance}

Examining the implementation requirements of the different solutions reveals that HABIT provides a notably scalable and performant solution. As listed in Table~\ref{tab:sizes}, the size of the frameworks based on our proposed framework HABIT increases with more refined H3 resolutions, but always achieve highly compressed representations of vessel traffic in all areas.
In contrast, GTI generates models of considerably larger size (by an order of magnitude), especially when compared to the volume of original AIS data. The GTI model size is highly sensitive to AIS data characteristics such as sparsity. The more confined KIEL dataset yielded a relatively smaller framework footprint, whereas the SAR dataset necessitated a larger, more complex structure.
Moreover, GTI required significant memory resources during the framework construction, and sometimes failed to conclude its computation necessary for identifying the best-fit parameters.

With regard to query response time, we measured average and maximum latency at  imputation tests under varying method configurations (Table~\ref{tab:latency}). The same gaps (duration 60 minutes) were tested  in all methods settings for the same area (KIEL, SAR). Notably, HABIT retains a subsecond latency including its simplification and reconstruction cost. As expected, for more refined H3 resolutions the latency increases. However, GTI incurs higher response times for better imputation results, and generally requires more time than HABIT, especially in the more complex SAR dataset which includes diverse vessel types and areas of different AIS coverage.
Overall, our proposed framework HABIT achieves higher scalability, being able to perform effectively at different maritime scenarios, while requiring reduced computational resources.




%% file: TEX/conclusion.tex
\section{Conclusions and Future Work}
\label{sec:conclusion}



This application paper presents a lightweight framework for vessel trajectory imputation based on an H3 graph that models area-specific movement patterns from batch AIS statistics. Through extensive fine-tuning, including path simplification and the incorporation of locally median positions during reconstruction, the framework achieves high geometric fidelity to true vessel paths. Empirical results show comparable, and in some cases superior, accuracy to a state-of-the-art imputation approach, while demonstrating strong scalability, sub-second query latency, and significantly lower memory requirements, even as data size and sparsity increase.

As a next step, we plan to expand our approach by incorporating features related to the vessel's state (e.g., draught) or voyage context (e.g., weather conditions, month of travel, destination port). In addition, we will evaluate transformer-based models \cite{10.1145/3555041.3589733, drapier2024enhancing, nadiri2025trajlearn} for the imputation task at scale, aiming to create a unified pipeline that improves data quality before generating area-specific analytics, such as density maps.

    

%% file: references.bib
@String { SIGMOD     = {{ACM} Intl. Conf. on Management of Data ({SIGMOD})}}

@String { PVLDB      = {VLDB Endowment ({PVLDB})}}

@String { SIGSPATIAL       = {{ACM Intl. Conf. on Advances in Geographic Information Systems (SIGSPATIAL)}}}

@String { TSAS       = {{ACM Trans. Spatial Algorithms Syst. (TSAS)}}}

@String { MDM       = {{IEEE Intl. Conf. on Mobile Data Management (MDM)}}}

@String { EDBT       = {Intl. Conf. on Extending Database Technology ({EDBT})}}

@String { KDD        = {{ACM} Intl. Conf. on Knowledge Discovery and Data Mining ({KDD})}}

@String { ICDE       = {{IEEE} Intl. Conf. on Data Engineering ({ICDE})}}

@String { SSTD       = {Intl. Symposium on Spatial and Temporal Data (SSTD)} }

@inproceedings{10.1145/3589132.3625620,
  author = {Isufaj, Keivin and Elshrif, Mohamed Mokhtar and Abbar, Sofiane and Mokbel, Mohamed},
  title = {{GTI: A Scalable Graph-Based Trajectory Imputation}},
  year = {2023},
  booktitle = SIGSPATIAL
}

@inproceedings{10.1145/3557915.3560942,
  author = {Elshrif, Mohamed M. and Isufaj, Keivin and Mokbel, Mohamed F.},
  title = {{Network-Less Trajectory Imputation}},
  year = {2022},
  booktitle = SIGSPATIAL
}

@article{10.14778/3632093.3632105,
  author = {Chen, Yile and Cong, Gao and Anda, Cuauhtemoc},
  title = {{TERI: An Effective Framework for Trajectory Recovery with Irregular Time Intervals}},
  year = {2023},
  journal = PVLDB
}

@article{10.14778/3773749.3773756,
      title={{MH-GIN}: Multi-scale Heterogeneous Graph-based Imputation Network for {AIS} Data}, 
      author={Hengyu Liu and Tianyi Li and Yuqiang He and Kristian Torp and Yushuai Li and Christian S. Jensen},
      year={2025},
      volume = {19},
      number = {2},
      journal = PVLDB
}

@inproceedings{10.1145/3555041.3589733,
  author = {Musleh, Mashaal and Mokbel, Mohamed},
  title = {{A Demonstration of KAMEL: A Scalable BERT-Based System for Trajectory Imputation}},
  year = {2023},
  booktitle = SIGMOD
}

@inproceedings{10591657,
  author = {Mohammed, Hayat Sultan and Nascimento, Mario A. and Barbosa, Denilson},
  title = {{Effective Trajectory Imputation Using Simple Probabilistic Language Models}},
  year = {2024},
  booktitle = MDM
}

@inproceedings{spiliopoulos2024patterns,
  title = {{Patterns of Life: Global Inventory for Maritime Mobility Patterns}},
  author = {Spiliopoulos, Giannis and Vodas, Marios and Grigoropoulos, Georgios and Bereta, Konstantina and Zissis, Dimitris},
  booktitle = EDBT,
  year = {2024}
}

@Article{rs15215080,
  title = {{The Big Picture: An Improved Method for Mapping Shipping Activities}},
  author = {Troupiotis-Kapeliaris, Alexandros and Zissis, Dimitris and Bereta, Konstantina and Vodas, Marios and Spiliopoulos, Giannis and Karantaidis, Giannis},
  journal = {Remote Sensing},
  volume = {15},
  number = {21},
  year = {2023}
}

@inproceedings{10.1145/3637528.3672086,
  author = {Zhang, Zhiwen and Fan, Zipei and Lv, Zewu and Song, Xuan and Shibasaki, Ryosuke},
  title = {{Long-Term Vessel Trajectory Imputation with Physics-Guided Diffusion Probabilistic Model}},
  year = {2024},
  booktitle = KDD
}

@article{WANG2025107279,
  title = {{TrajDiff: A Method for Vessel Trajectory Imputation Utilizing Resampled Conditional Diffusion Models}},
  journal = {Results in Engineering},
  volume = {28},
  year = {2025},
  author = {Wentao Wang and Wei Xiong and Luo Chen and Hao Chen}
}

@inproceedings{chen2023rntrajrec,
  title = {{RNTrajRec: Road Network Enhanced Trajectory Recovery with Spatial-Temporal Transformer}},
  author = {Chen, Yuqi and Zhang, Hanyuan and Sun, Weiwei and Zheng, Baihua},
  booktitle = ICDE,
  year = {2023}
}

@inproceedings{DBLP:conf/edbt/0002ZST18,
  author = {Chatzikokolakis, Konstantinos and Zissis, Dimitrios and Spiliopoulos, Giannis and Tserpes, Konstantinos},
  title = {{Mining Vessel Trajectory Data for Patterns of Search and Rescue}},
  booktitle = {{Big Mobility Data Analytics Workshop (BMDA)}},
  year = {2018}
}

@inproceedings{10.1145/3609956.3609961,
  author = {Magnussen, B\'{u}gvi Benjamin and Bl\"{a}ser, Nikolaj and Lu, Hua},
  title = {{DAISTIN: A Data-Driven AIS Trajectory Interpolation Method}},
  year = {2023},
  booktitle = SSTD
}

@article{10.1007/s10707-022-00475-0,
  author = {Fikioris, Giannis and Patroumpas, Kostas and Artikis, Alexander and Pitsikalis, Manolis and Paliouras, Georgios},
  title = {{Optimizing Vessel Trajectory Compression for Maritime Situational Awareness}},
  year = {2022},
  volume = {27},
  number = {3},
  journal = {Geoinformatica}
}

@article{10.1007/s10707-016-0266-x,
  author = {Patroumpas, Kostas and Alevizos, Elias and Artikis, Alexander and Vodas, Marios and Pelekis, Nikos and Theodoridis, Yannis},
  title = {{Online Event Recognition from Moving Vessel Trajectories}},
  year = {2017},
  volume = {21},
  number = {2},
  journal = {Geoinformatica}
}

@article{10.1145/3656470,
  author = {Musleh, Mashaal and Mokbel, Mohamed F.},
  title = {{Let's Speak Trajectories: A Vision to Use NLP Models for Trajectory Analysis Tasks}},
  year = {2024},
  journal = TSAS
}

@inproceedings{berndt1994using,
  title = {{Using Dynamic Time Warping to Find Patterns in Time Series}},
  author = {Berndt, Donald J. and Clifford, James},
  booktitle = KDD,
  year = {1994}
}

@inproceedings{visvalingam1990douglas,
  title = {{The Douglas--Peucker Algorithm for Line Simplification: Re-Evaluation through Visualization}},
  author = {Visvalingam, Mahes and Whyatt, J. Duncan},
  booktitle = {Computer Graphics Forum},
  volume = {9},
  number = {3},
  year = {1990}
}

@article{yang2019big,
  title = {{How Big Data Enriches Maritime Research: A Critical Review of Automatic Identification System (AIS) Data Applications}},
  author = {Yang, Dong and Wu, Lingxiao and Wang, Shuaian and Jia, Haiying and Li, Kevin X.},
  journal = {Transport Reviews},
   year = {2019}
}

@article{nadiri2025trajlearn,
  title = {{TrajLearn: Trajectory Prediction Learning Using Deep Generative Models}},
  author = {Nadiri, Amirhossein and Li, Jing and Faraji, Ali and Abuoda, Ghadeer and Papagelis, Manos},
  journal = TSAS,
  volume = {11},
  number = {3},
  year = {2025}
}

@article{drapier2024enhancing,
  title = {{Enhancing Maritime Trajectory Forecasting via H3 Index and Causal Language Modelling (CLM)}},
  author = {Drapier, Nicolas and Chetouani, Aladine and Chateigner, Aur{\'e}lien},
  journal = {arXiv:2405.09596},
  year = {2024}
}

@inproceedings{troupiotis2025dynamic,
  author = {Troupiotis-Kapeliaris, A. and Grigoropoulos, G. and Vodas, M. and Bereta, K.},
  title = {{Dynamic Weather-Resilient Vessel Routing Using Big AIS Data [Industry]}},
  booktitle = SIGSPATIAL,
  year = {2025}
}

@inproceedings{WuTTZZS25,
  author       = {Song Wu and
                  Kristian Torp and
                  Alexandros Troupiotis{-}Kapeliaris and
                  Dimitris Zissis and
                  Esteban Zim{\'{a}}nyi and
                  Mahmoud Attia Sakr},
  title        = {Effective Ship Trajectory Imputation with Multiple Coastal Cameras},
  booktitle    =  MDM,
   year         = {2025}
 }
